\documentclass[11pt,a4paper]{article}
\pdfoutput=1
\usepackage{empheq}
\usepackage{cases}
\usepackage{a4wide}
\usepackage{amsfonts}
\usepackage{amssymb}
\usepackage{amsmath,bm}
\usepackage{amsmath}
\usepackage{graphicx}
\usepackage{mathtools}
\usepackage{booktabs}
\usepackage{array}
\usepackage{color}
\usepackage{ulem}
\usepackage[small,bf]{caption}
\usepackage{cite}
\usepackage[usenames,dvipsnames]{xcolor}
\usepackage{subfigure}
\usepackage{verbatim}

\allowdisplaybreaks

\numberwithin{equation}{section}

\newcounter{mysubequation}[equation]

\DeclarePairedDelimiterX\braket[2]{\langle}{\rangle}{#1 \delimsize\vert #2}

\begin{document}
\begin{titlepage}

\begin{center}
{
\bf\LARGE 
Impacts of ALP on the Constraints of Dark Photon 
}
\\[8mm]
Chuan-Ren~Chen\footnote[1]{crchen@ntnu.edu.tw},~Yuan-Feng Hsieh\footnote[2]{61041026S@ntnu.edu.tw},~and Chrisna~Setyo~Nugroho\footnote[3]{setyo13nugros@ntnu.edu.tw}  
\\[1mm]
\end{center}
\vspace*{0.50cm}

\centerline{\it Department of Physics, National Taiwan Normal University, Taipei 116, Taiwan}
\vspace*{1.20cm}

\begin{abstract}
\noindent

Dark sector may exist and interact with Standard Model (SM)
through the $U(1)$ kinetic mixing. Through this portal-type interaction, dark photon from dark sector couples to SM 
fermions, and may explain the discrepancy between experimental data and SM calculations on muon anomalous magnetic moment,
muon $g-2$. However, current searches for dark photon impose stringent constraints on the mixing parameter $\varepsilon$
for various dark photon masses, excluding the favorite parameter space for muon $g-2$. In this paper, we study the
case where a global $U(1)$ in dark sector is spontaneously
broken, resulting a light pseudo-Goldstone, axion-like particle (ALP) $a$, which couples to dark photon and SM
photon, $g_{a\gamma\gamma'}$. Through this interaction, dark
photon may decay into photon and ALP when this channel is
kinematically allowed. As a result, the experimental
constraints on dark photon change significantly, and dark photon is able to explain the muon $g-2$ anomaly when its mass
is heavier than $10$ GeV.       

\end{abstract}

\end{titlepage}
\setcounter{footnote}{0}

\section{Introduction}

Besides the discovery of Higgs boson, the Large Hadron Collider (LHC) is built for new physics signals. However, the null results after Run II at the LHC indicate that new physics may interact  extremely weak with Standard Model (SM). The existence of dark sector that interacts with SM through portals naturally realizes this possibility. Depending on the structure of dark sector, the portal can be realized via scalars, sterile neutrinos or vectors.  
The simplest mechanism for vector portal is given by a kinetic mixing term between two $U(1)$ gauged symmetries from dark sector and SM, respectively. As a result, a new gauge boson $\gamma'$, referred to dark photon, interacts with the SM electromagnetic current $J_{em}^\mu$ as $\varepsilon A'_\mu J_{em}^\mu$, where $A'_\mu$ is the dark photon field and  $\varepsilon$ is the mixing parameter \cite{Holdom:1985ag,Okun:1982xi,Fayet:1980rr,Georgi:1983sy}. 
     

Several proposals have been
made to detect dark photon, ranging from laboratory experiments to
astronomical observations\cite{Workman:2022ynf, Fabbrichesi:2020wbt,Caputo:2021eaa,Hook:2021ous,Carenza:2023qxh,Kalashev:2018bra}, and the $\varepsilon$ is constrained to be tiny and less than about $10^{-3}$ for dark photon mass below about $100$ GeV. 
On the other hand, the interactions between SM fermions and dark photon introduce additional one-loop diagrams to anomalous magnetic moments ($g-2$). The long-standing tension between SM calculation and experimental measurements by BNL \cite{Muong-2:2006rrc} and FermiLab~\cite{Muong-2:2023cdq} on muon $g-2$ reaches $5\sigma$ level. It has been shown that such anomaly can be explained by a sub-GeV $\gamma'$~\cite{Pospelov:2008zw}. However, the collider searches of $\gamma'$ exclude this parameter space favorited by muon $g-2$.

Note that the constraints from collider searches mainly come from the null excesses on signals of $e^+e^-$  or $\mu^+\mu^-$ pairs from $\gamma'$ decays. If the structure of dark sector allows additional decay channels other than SM fermion final states, the total event rates of $e^+e^-$ and $\mu^+\mu^-$ would be significantly changed. 
In this paper, we propose the existence of a pseudo-Goldstone boson when a global $U(1)$ in dark sector is broken, which is referred to axion-like particle (ALP). Similar to the QCD axion coupling to two photons, ALP interactes with dark photons and photons~\cite{Kaneta:2016wvf}. As a result, dark photon will decay into  ALP and  photon if ALP is lighter than dark photon\footnote{For extremely light ALP, it could decay into two photons which is constrained by the survey of infrared photon spectrum by JWST~\cite{Roy:2023omw}.}. 
We will see later that the decay channel of $\gamma'\to a \gamma$, where $a$ and $\gamma$ are ALP and photon, respectively, could dominate the the decay channels for $\gamma'$ heavier about $100$ MeV and rescue the excluded parameter space for muon $g-2$.

The  paper is organized as follows: In section~\ref{sec:model}, we
discuss the model containing the SM and dark sector with dark photon acting as a messenger between them. We
provide the current experimental limits on dark photon coupling in Section~\ref{sec:constraints}. We recast the recent experimental constraints in the presence of new interaction in Section~\ref{sec:newlimits} and further discuss its consequences there.
Our summary  is presented in Section~\ref{sec:Summary}.

\section{The Model}
\label{sec:model}

We focus on the dark sector where both global and gauged $U(1)$ symmetries are implemented. After the symmetries are spontaneously broken, a pseudo-Nambu-Goldstone boson dubbed axion-like particle (ALP), and a gauge boson dubbed dark photon are generated~\cite{Kaneta:2016wvf,Lee:2018yak}. The relevant Lagrangian are given as~\cite{Kaneta:2016wvf}
\begin{equation}
\label{eq:model}
{\cal L} \supset  \frac{1}{2}m_{\gamma'}^2A'_\mu A'^\mu+\frac{1}{2}\varepsilon F_{\mu\nu}F'^{\mu\nu} -\frac{1}{2}g_{a\gamma\gamma'}a F_{\mu\nu} \tilde{F'}^{\mu\nu},
\end{equation}
where $A'_\mu$ is dark photon field, $F_{\mu\nu}$ and $F'_{\mu\nu}$ are field strength of photon and dark photon, respectively; $\tilde{F}'_{\mu\nu}$ is the dual tensor of $F'_{\mu\nu}$, while $a$ represents the ALP. Because of the mixing term of photon and dark photon, the couplings between dark photon and fermions of the SM are introduced. As a result, the dark photons can be produced in the QED processes by simply replacing non-propagating photons by dark photons.  
Therefore, the production mechanism includes bremsstrahlung, annihilation, meson decays and Drell-Yan process.  
Regarding its decay, 
the dark photon $\gamma'$ will decay into SM fermions if its mass is heavy enough. The partial decay width of dark photon into a pair of charged leptons $\ell$ is 
\begin{equation}
\Gamma (\gamma '\to \ell^+ \ell^-) = \frac{\varepsilon ^{2} e^{2}}{12 \pi} m_{\gamma'}
 (1+2 \frac{m_{\ell}^2}{m_{\gamma'}^2} )\sqrt{1-4\frac{m_\ell^2}{m_{\gamma'}^2}}, 
\end{equation}
where $e$ and $m_\ell$ are the electric charge and mass of charged lepton $\ell$.  For quark channels of dark photon decay, the final state is represented by the hadrons. The partial decay width is therefore presented as
\begin{equation}
\Gamma (\gamma '\to hadrons) = \Gamma (\gamma '\to \mu^+ \mu^-)  R_t,
\end{equation}
where $R_t$ is defined as the ratio  $R_t\equiv\sigma(e^+e^-\to hadrons)/\sigma(e^+e^-\to\mu^+\mu^-)$, with $\sigma$ being the cross section at the $e^+e^-$ collider at a certain energy. For the numerical results shown later, we read out the values of  $R_t$ from PDG~\cite{pdg-R}. In addition, the dark photon may decay into an ALP and a photon due to the interaction $-\frac{1}{2}g_{a\gamma\gamma'}a F_{\mu\nu} \tilde{F'}^{\mu\nu}$ if it is kinematically allowed. In this case, we obtain the partial decay width as 
\begin{equation}
\Gamma (\gamma '\to a \gamma) = \frac{g_{a \gamma \gamma'}^{2}}{96 \pi} m_{\gamma'}^3
 (1- \frac{m_{a}^2}{m_{\gamma'}^2})^3,
\end{equation}
where $m_a$ is the mass of ALP.  Through this paper, we focus on the scenario where $m_{\gamma'}\gg m_a$, leading to a simple result $\Gamma (\gamma '\to a \gamma)\propto m_{\gamma'}^3$. 
The size of photon-dark-photon mixing parameter $\varepsilon$ and strength of $g_{a\gamma\gamma'}$ depend on the details of the UV setup of the model. Here, we treat them as parameters for simplicity in our phenomenology study. We notice that the constraints on $\varepsilon$ has been studied in details in the literature, see e.g.~\cite{Fabbrichesi:2020wbt} and references therein, if the $g_{a\gamma\gamma'}\ll\varepsilon$. However, when $g_{a\gamma\gamma'}$ is of the same order of $\varepsilon$, the current constraints on $\varepsilon$ change significantly and  searching strategy of dark photon should be revisited, as we will discuss later in Section~\ref{sec:newlimits}.

%
\begin{figure}
	\centering
	\includegraphics[width=0.47\textwidth]{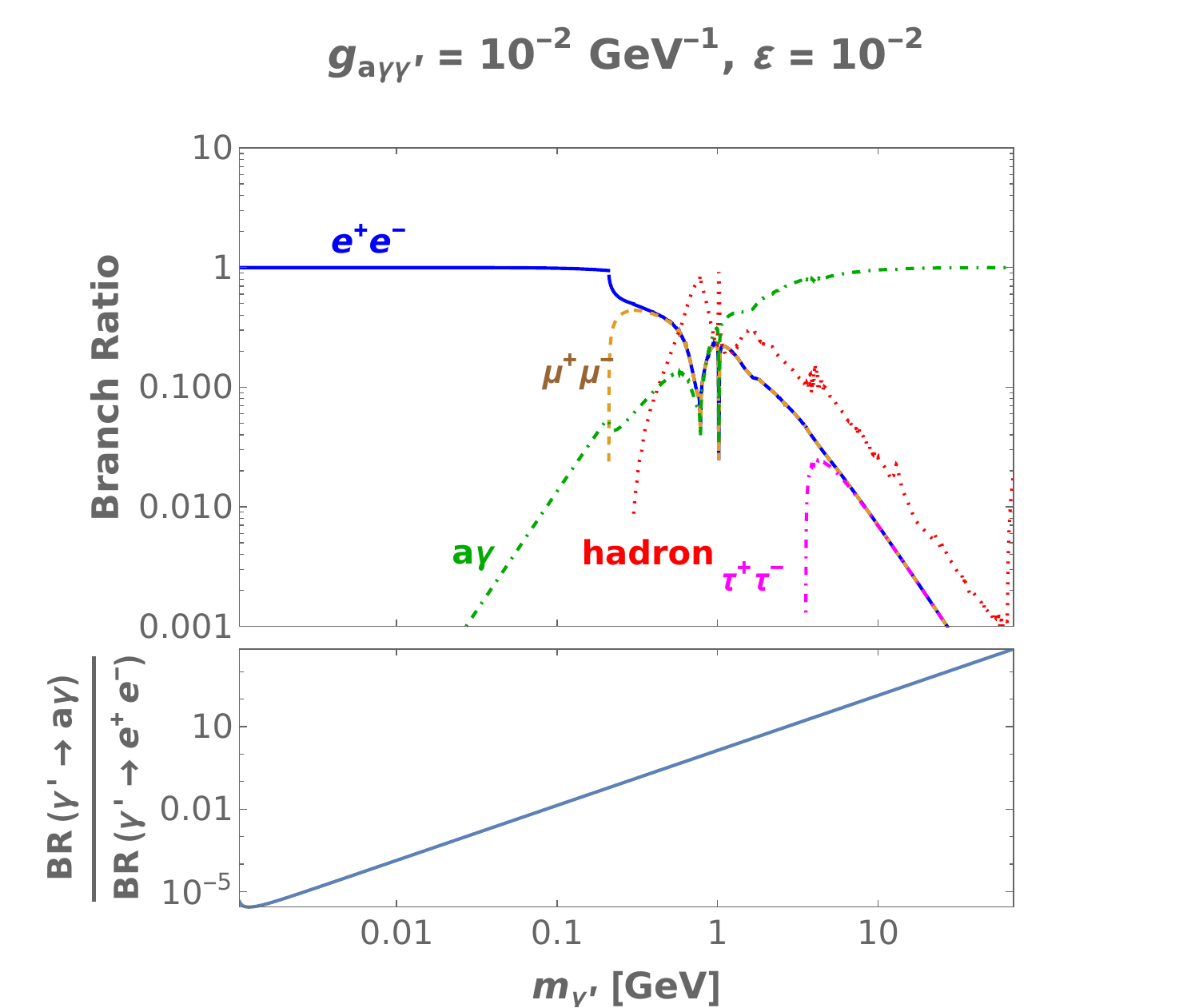}
	\includegraphics[width=0.47 \textwidth]{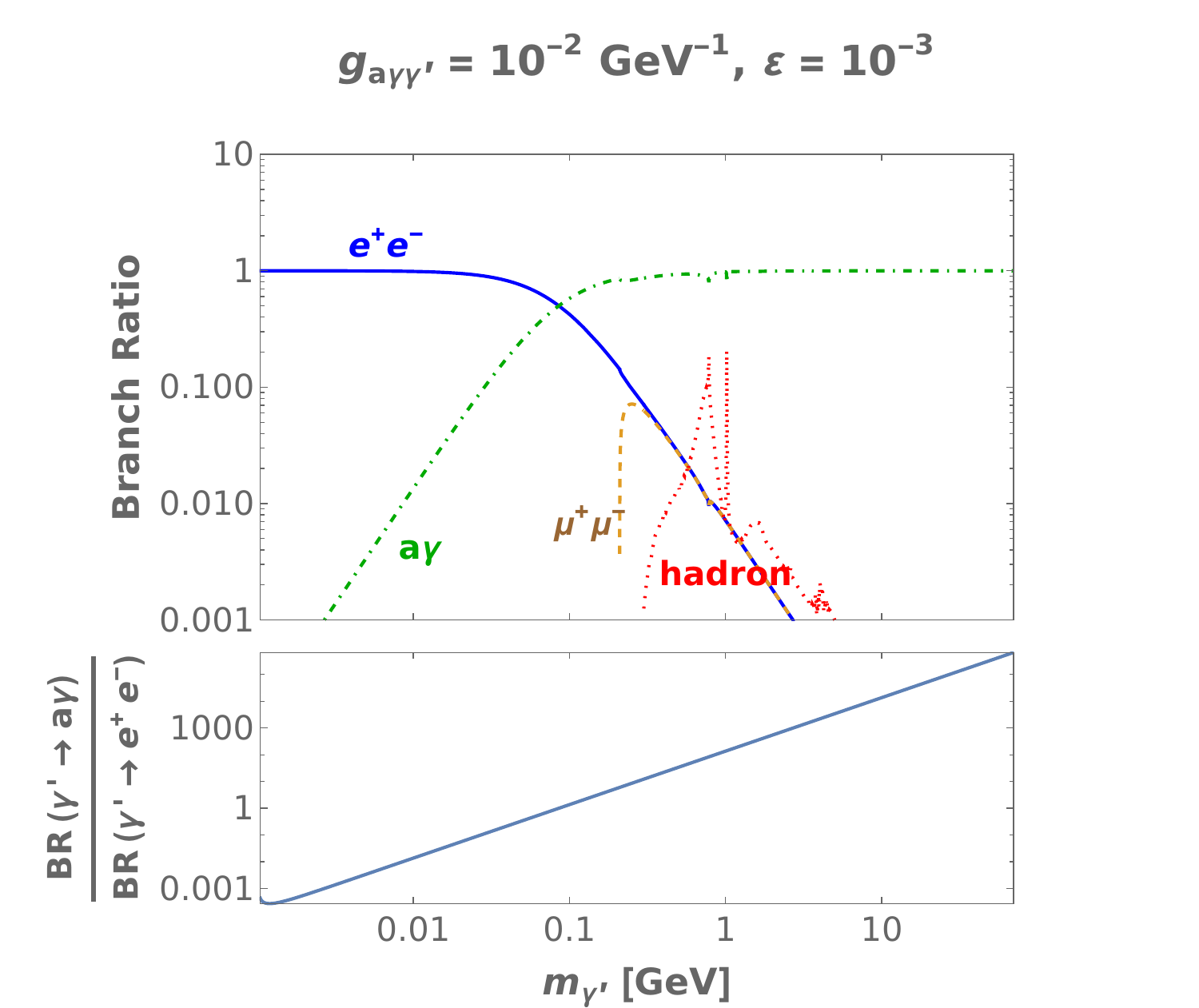}
	\caption{The decay branching ratios of dark photon. Two benchmark values, $10^{-2}$ and $10^{-3}$ are taken for mixing parameter $\varepsilon$, while effective dark photon-ALP-photon coupling $g_{a\gamma\gamma'}$ is fixed to be $10^{-2}~\rm{GeV^{-1}}$.}	
	\label{fig:brs}
\end{figure}

Fig.~\ref{fig:brs} shows the decay branching ratios of a dark photon and the ratio of branching ratio $\gamma'\to e^+e^-$ to  branching ratio $\gamma'\to a\gamma$ for its mass $m_{\gamma'}> 1~\rm{MeV} $, where  benchmark values $\varepsilon =10^{-2}$ (left) and $\varepsilon =10^{-3}$ (right) are taken, while $g_{a\gamma\gamma'}$ is fixed to be $10^{-2}~\rm{GeV^{-1}}$.
 When the value of $g_{a\gamma\gamma'}$ is as large as the mixing parameter $\varepsilon$, see the left pannel, dark photon will mainly decay into $\gamma$ and ALP when $m_{\gamma'}$ is larger than $1$ GeV, while hadronic modes are subdominant. However, for a light dark photon lighter than about $200$ MeV, $e^+e^-$ mode is the main channel. The branching ratio of $\gamma'\to\mu^+\mu^-$ is also the same as $e^+e^-$ mode and becomes the main decay channel up to $m_\gamma' \simeq 700~\rm{MeV}$. For the window of dark photon mass between $700$ MeV and $1$ GeV, hadronic decay modes of dark photon is the leading channel.     
  For  $\varepsilon = 10^{-3}$ and $g_{a\gamma\gamma'}=10^{-2}~\rm{GeV^{-1}}$, $e^+e^-$ mode is the dominant decay channel till $m_{\gamma'} \simeq 100~\rm{MeV}$.  Beyond $100$ MeV, the $a\gamma$ channel takes over.

It is worth mentioning that
the product of $\varepsilon g_{a\gamma\gamma'}$ can be bounded using the light shining through the wall (LSW) experiments~\cite{Inada:2013tx,Lee:2022myb,Ismail:2022ukp} in the case of very light dark photon such that the decay channel to $e^+e^-$ is absent. The reason can be easily understood since the converting probability of photon to dark photon is proportional to $\varepsilon^2$ while the decay probability inside the detector is approximately proportional to decay width which is proportional to $g_{a\gamma\gamma'}^2$. The conclusion is that, as the dark photon is of order of $O(10^4)$ eV, $\varepsilon g_{a\gamma\gamma'}$ needs to be smaller than about $ 10^{-7}~\rm{GeV^{-1}}$. The limit becomes weaker when mass of dark photon is even lighter. For $m_{\gamma'}\lesssim 10^{-1}$ eV, $\varepsilon g_{a\gamma\gamma'}$ can be as large as $10^{-3}~\rm{GeV^{-1}}$.  Note that since we focus on the case that $m_{\gamma'}$ is heavier than $1$ MeV, the LSW experiment has no sensitivity on dark photon search.  
For the coupling strength $g_{a\gamma\gamma'}$ along, B-factories, e.g. BaBar, can impose stringent constraints  through the search for mono-photon signal~\cite{BaBar:2014zli}. Due to limitation of the electron-beam energy,  B-factories can only examine dark photon up to $10$ GeV, and it is shown that  the region of $g_{a\gamma\gamma'}\gtrsim 3\times 10^{-3}~\rm{GeV^{-1}}$ is excluded~\cite{deNiverville:2018hrc}.

\section{Experimental Limits}
\label{sec:constraints}

\begin{figure}
	\centering
	\includegraphics[width=0.6\textwidth]{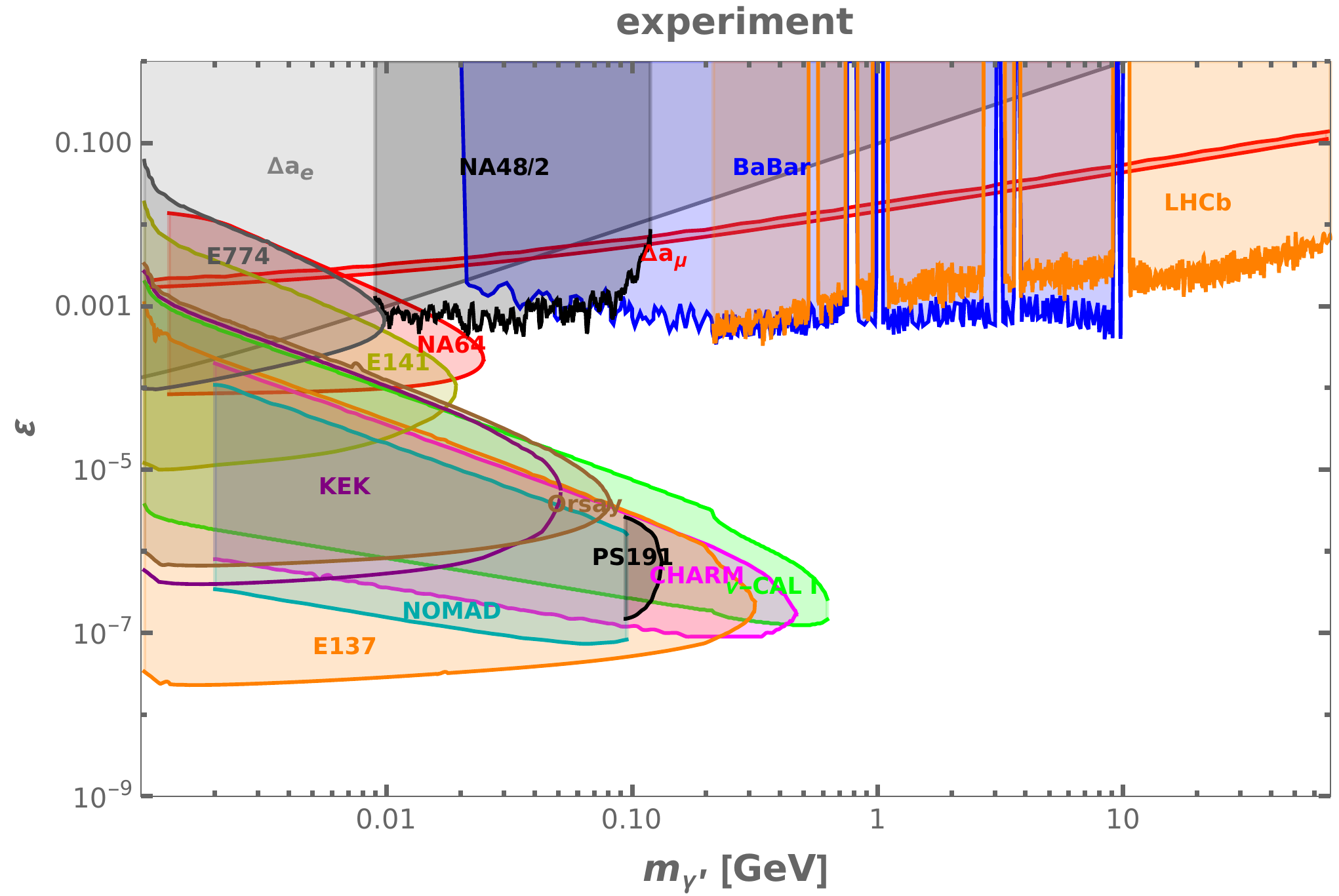}
	\caption{The current constraints of mixing parameter $\varepsilon$ for $m_{\gamma'}$ in the region of $1$ MeV to $100$ GeV. The shaded area is excluded~\cite{Ilten:2018crw}. The parameter space excluded by electron anomalous magnetic moment measurement ($\Delta a_e$) and the one favored by muon ($\Delta a_\mu$) are indicated by gray region and red band, respectively. }	
	\label{fig:currentlimit}
\end{figure}

Many experiments put efforts to search for dark photon, including colliders, fixed targets, and beam dumps. 
We focus on the one that searches for signals of electron and positron pairs from dark photon decay, since it is an inevitable decay channel of dark photon. 
We don't include invisible decay channel because this channel depends on dark matter assignment in dark sector. A summary of the current constraints on $\varepsilon$-$m_{\gamma'}$ plane  is displayed in Fig.~\ref{fig:currentlimit} which is adopted from Fig. 4 of~\cite{Ilten:2018crw}. 

For dark photon mass considered in this paper, $1\,\text{MeV} < m_{\gamma^{'}} < 100\, \text{GeV}$, several experiments has
been carried out to probe dark photon in the laboratories. In
collider frontier, Babar collaboration puts stringent limit on
dark photon parameter space based on visible search  of $\gamma'$ decay into $e^{+} e^{-}$ and $\mu^{+} \mu^{-}$~\cite{BaBar:2014zli}. Furthermore, NA48/2 experiment has performed dark
photon search from pion decay $\pi^{0} \rightarrow \gamma' \gamma$ followed by $\gamma' \rightarrow e^{+} e^{-}$~\cite{NA482:2015wmo}. In addition, A1 and APEX collaborations
employ electron bremsstrahlung in fixed target experiment $eZ \rightarrow eZ \gamma'$ with $e^{+} e^{-}$ as final state of $\gamma'$ decay~\cite{Merkel:2014avp,APEX:2011dww}. Moreover, KLOE experiment has been looking for dark photon from $\phi \rightarrow \gamma' \eta$ followed by $\gamma' \rightarrow e^{+} e^{-}$ as well as $e^{+} e^{-} \rightarrow \gamma' \gamma$ with $\gamma' \rightarrow \pi^{+} \pi^{-}$ decay mode \cite{KLOE-2:2011hhj,KLOE-2:2016ydq}. Finally, the LHCb collaboration has performed dark photon search using $\gamma' \rightarrow \mu^{+} \mu^{-}$ decay mode from $pp \rightarrow \gamma'$ production channel \cite{LHCb:2017trq}.

On the other hand, beam dump experiments have put stringent
limits on dark photon search, especially in sub-Gev dark
photon mass. In both electron and proton beam dumps, dark
photon search is inferred from its decay to $e^{+} e^{-}$
final state. For electron beam dump, several experiments such
as Orsay, E774, E137, E141, KEK, and NA64 has placed strict limits on dark photon~\cite{Davier:1989wz,Bross:1989mp,Bjorken:1988as,Riordan:1987aw,Konaka:1986cb,NA64:2019auh}. Moreover, proton beam dump experiments
like PS191, NOMAD, CHARM, $\nu-$CAL I also exclude dark photon parameter space~\cite{Bernardi:1985ny,NOMAD:2001eyx,CHARM:1985anb,Blumlein:1990ay,Blumlein:1991xh}. Furthermore, anomalous magnetic moment of both electron and muon put additional limits on dark photon parameter space~\cite{Pospelov:2008zw,Endo:2012hp,Muong-2:2021ojo,Aoyama:2020ynm,Bodas:2021fsy}. All these limits are
presented in Fig.\ref{fig:currentlimit} with exception of A1,
APEX, and KLOE limits since they are already covered by BaBar
experiment.

\section{New Limits on Photon-Dark Photon Mixing}
\label{sec:newlimits}

In all experiments mentioned above, the searches of dark
photon focus on detecting electron and positron pairs from
dark photon decay. The number of event ($N$) of these pairs in
dark photon experiments can be described by
\begin{eqnarray}
\label{eq:event}
N = \cal{L} \cdot \sigma_{\text{Prod}}(\gamma^{'}) \cdot \text{BR}(\gamma^{'}) \cdot \eta (\tau_{\gamma^{'}})\,.
\end{eqnarray}
Here, $\cal{L}\,, \sigma_{\text{Prod}}(\gamma^{'})\,, \text{BR}(\gamma^{'})\,, \text{and}\, \eta (\tau_{\gamma^{'}}) $ stand for the integrated luminosity, dark photon production cross
section, branching ratio of dark photon decay into $e^+e^-$, and detector efficiency,
respectively. In the presence of new photon-dark photon-axion 
interaction given by the last term of Eq.\eqref{eq:model}, one
expects that, for $m_{\gamma^{'}} > m_{a}$ considered in this work, dark
photon could decay into the axion and photon. As a result, 
the $e^+e^-$ branching ratio and lifetime of $\gamma'$ will be different while leaving its production cross section intact. Therefore, limit on photon-dark photon mixing would be modified and should be revisited. 

In term of the observed event number of $e^+e^-$, one has the relation 
\begin{align}
\label{eq:Revent}
\cal{L}^{\text{New}} \cdot \sigma^{\text{New}}_{\text{Prod}}(\gamma^{'}) \cdot \text{BR}^{\text{New}}(\gamma^{'}) \cdot \eta^{\text{New}} (\tau_{\gamma^{'}})= \cal{L}^{\text{Old}} \cdot \sigma^{\text{Old}}_{\text{Prod}}(\gamma^{'}) \cdot \text{BR}^{\text{Old}}(\gamma^{'}) \cdot \eta^{\text{Old}} (\tau_{\gamma^{'}}) ,
\end{align} 
where the superscript New (Old) denotes the photon-dark photon
interaction with (without) axion. Since the integrated
luminosity in an experiment is independent of the interaction in dark sector, one obtains $\cal{L}^{\text{New}}=\cal{L}^{\text{Old}}  $. Furthermore, for $m_{\gamma^{'}} > m_{a}$, the ratio of dark photon production cross section is 
\begin{align}
\label{eq:RCrosSec}
\frac{\sigma^{\text{New}}_{\text{Prod}}(\gamma^{'})}{\sigma^{\text{Old}}_{\text{Prod}}(\gamma^{'})} \approx \frac{\varepsilon^{2}_{\text{New}}}{\varepsilon^{2}_{\text{Old}}}\,,
\end{align}
where we have used the fact that the dark photon production cross section scales as $\varepsilon^{2}$ at colliders, and the dark photon from meson decays is much less than the dominant decay channels. Thus, we can rewrite the number of event equation as
\begin{align}
\label{eq:ReventNew}
\varepsilon^{2}_{\text{New}} \cdot \text{BR}^{\text{New}}(\gamma^{'}) \cdot \eta^{\text{New}} (\tau_{\gamma^{'}}) \approx
 \varepsilon^{2}_{\text{Old}} \cdot \text{BR}^{\text{Old}}(\gamma^{'}) \cdot \eta^{\text{Old}} (\tau_{\gamma^{'}}) \,.
\end{align}
In addition, the explicit expression of the new branching ratio $\text{BR}^{\text{New}}(\gamma^{'})$ and the old one $\text{BR}^{\text{Old}}(\gamma^{'})$ is given by
\begin{align}
\label{eq:BR}
\text{BR}^{\text{New}}(\gamma^{'}) &= \text{BR}^{\text{Old}}(\gamma^{'}) \left( 1+\frac{\Gamma(\gamma' \to a\gamma)}{\Gamma^{\text{Old}}_{\text{tot}}} \right)^{-1}\,,
\end{align}
where $\Gamma^{\text{Old}}_{\text{tot}}$ stands for the total decay width of dark photon without including ALP and $\Gamma(\gamma' \to a\gamma)$ is the partial decay width of dark photon decay into ALP plus a photon. 

Finally, the detection efficiency $\eta$ depends on the lifetime of dark photon, and varies according to the details of the experimental setups. 
The visible search for dark photon from BaBar relies on the prompt decay of dark photon into $e^+e^-$ and $\mu^+\mu^-$ after being produced. 
Therefore,  $\eta^{\text{New}}(\tau_{\gamma^{'}})$ should be equal to $\eta^{\text{Old}}(\tau_{\gamma^{'}})$. 
In beam dump experiments, the situation is more complicated.
Following the analysis in~\cite{Ilten:2018crw}, it is possible to construct the effective
proper-time in a given decay region $\left[\tilde{t}_{0}, \tilde{t}_{1} \right]$. In this case, $\tilde{t}_{0}$ is the time dark photon spend inside the beam dump shield, and $\tilde{t}_{1}$ denotes the required time for dark photon to
reach the detector from its production point. These two quantities are related by
\begin{align}
\label{eq:t0t1}
\tilde{t}_{1} = \tilde{t}_{0} \left(1 + \frac{L_{\text{dec}}}{L_{\text{sh}}} \right)\,,
\end{align}
where $L_{\text{dec}}$ and $L_{\text{sh}}$ stand for the
length of decay volume and the length of the
shield, respectively. Therefore, the corresponding efficiency ratio becomes
\begin{align}
\label{eq:REtaBD}
\frac{\eta^{\text{New}}(\tau_{\gamma^{'}})}{\eta^{\text{Old}}(\tau_{\gamma^{'}})} = \frac{e^{-\tilde{t}_{0}/\tau^{\text{New}}_{\gamma^{'}}} -  e^{-\tilde{t}_{1}/\tau^{\text{New}}_{\gamma^{'}}}}{e^{-\tilde{t}_{0}/\tau^{\text{Old}}_{\gamma^{'}}} -  e^{-\tilde{t}_{1}/\tau^{\text{Old}}_{\gamma^{'}}}}\,.
\end{align}
To obtain the value of $\tilde{t}_{0}$ at each dark photon mass $m_{\gamma^{'}}$ inside the interval limits of $\left[\varepsilon_{\text{min}}, \varepsilon_{\text{max}} \right]$, one needs to solve the following equation~\cite{Ilten:2018crw}
\begin{align}
\label{eq:t0Solve}
\varepsilon^{2}_{\text{max}} \, \eta\left[\tau_{\gamma^{'}} (\varepsilon^{2}_{\text{max}}) \right] = \varepsilon^{2}_{\text{min}} \, \eta\left[\tau_{\gamma^{'}} (\varepsilon^{2}_{\text{min}}) \right]\,,
\end{align}
which stems from the fact that the observed numbers of signals for setting the bound are the same.

\begin{figure}
	\centering
	\includegraphics[width=0.45\textwidth]{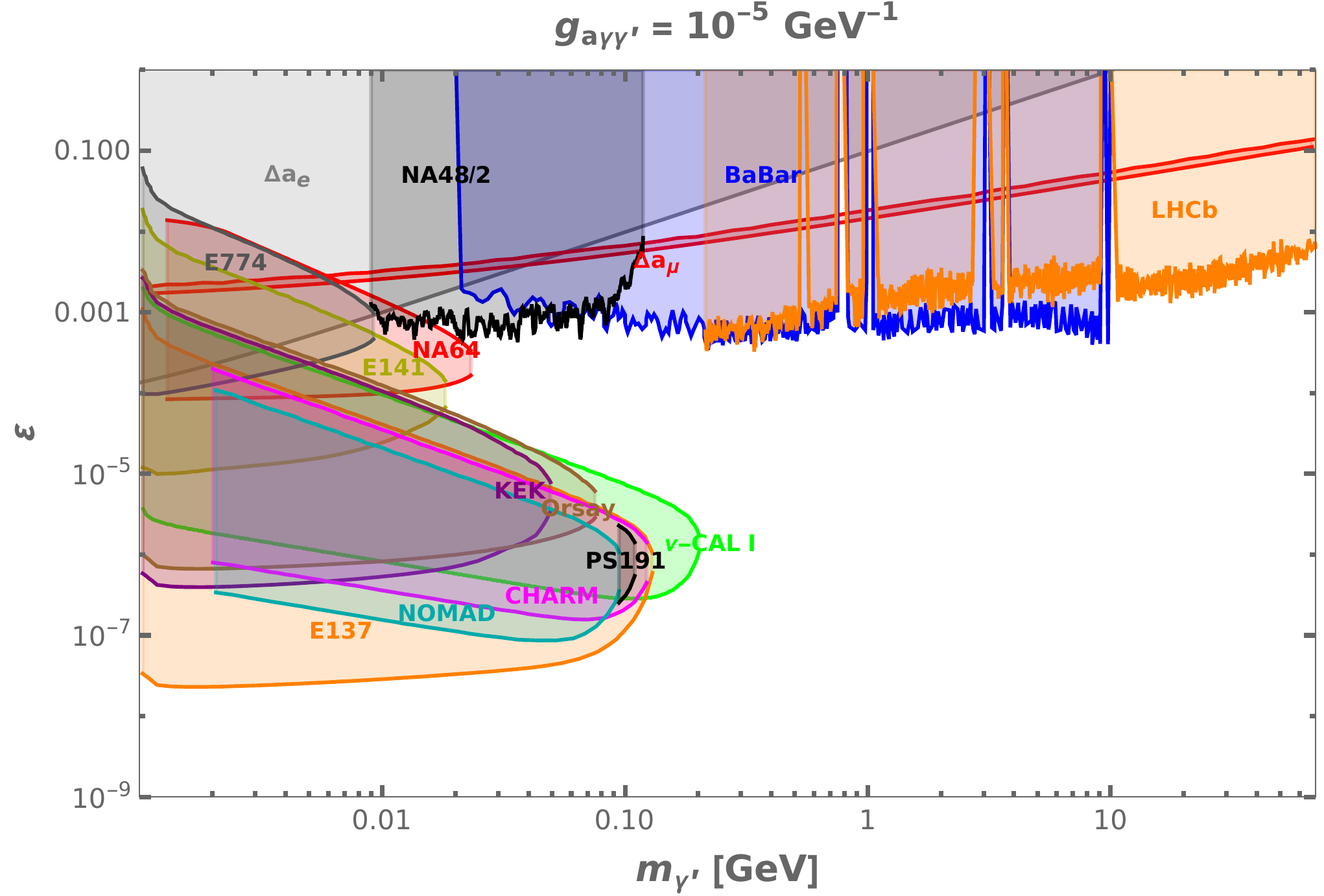}
	\includegraphics[width=0.45\textwidth]{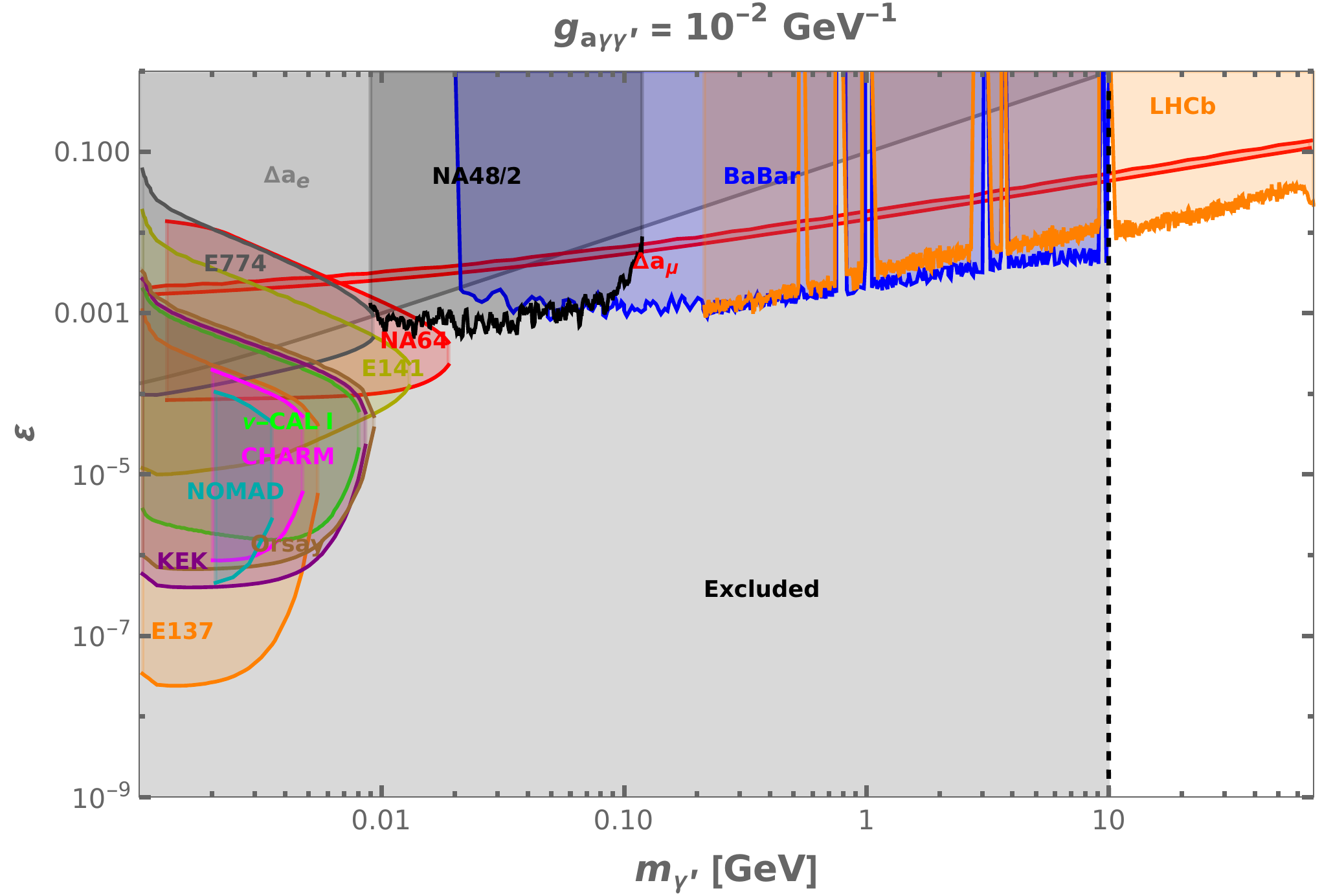}
	\includegraphics[width=0.45\textwidth]{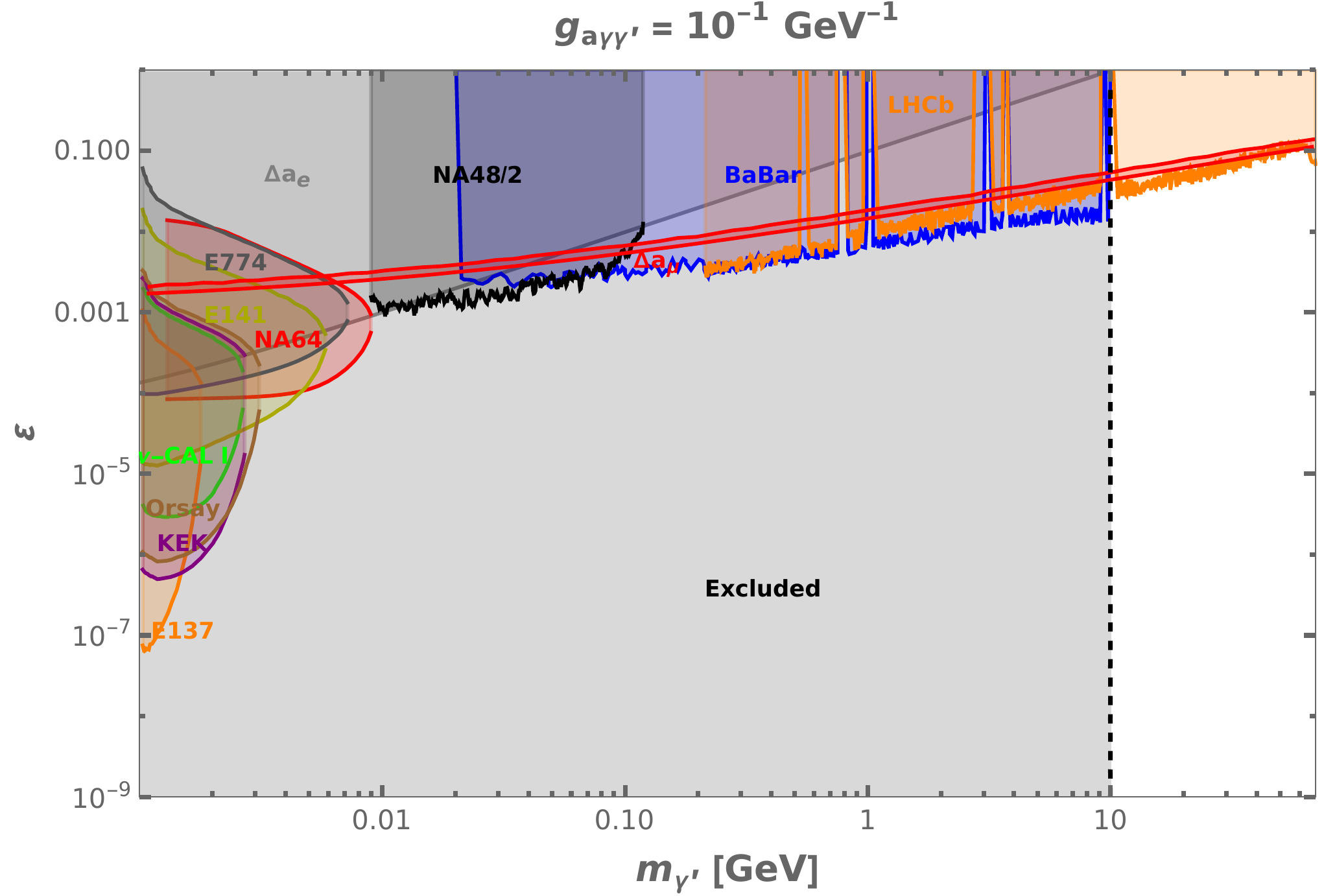}
	\includegraphics[width=0.45\textwidth]{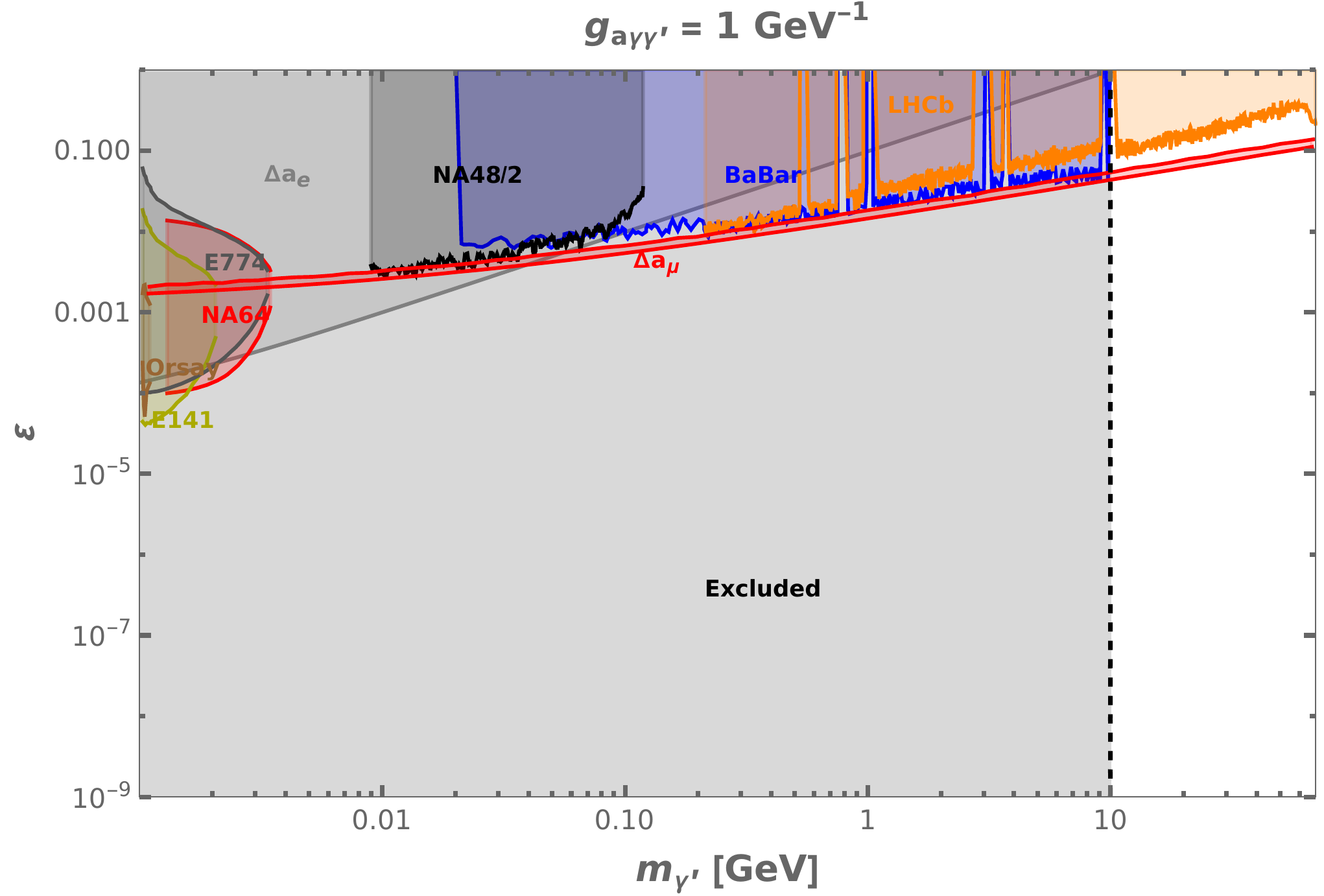}
	\caption{The limits on $\varepsilon$-$m_{\gamma'}$ plane in the presence of ALP-photon-dark photon interaction. The light gray region below 10 GeV for $g_{a\gamma\gamma'} \geq 10^{-4}$ is excluded~\cite{deNiverville:2018hrc,Jodlowski:2023yne}.}	
	\label{fig:limit}
\end{figure}

The impacts of ALP in dark sector on dark photon searches are shown in Fig.~\ref{fig:limit}. When the decay channel $\gamma'\to a\gamma$ is opened,  even though the coupling strength $g_{a\gamma\gamma'}=10^{-5}~\rm{GeV}^{-1}$ is small, its effect shows up for mass of dark photon around $0.1$ GeV in some beam dump experiments covering the region of $10^{-3}\gtrsim\varepsilon\gtrsim 10^{-8}$, e.g. E137 and $\nu$-CAL I, see upper-left panel of Fig.\ref{fig:limit}. This is because these experiments have long decay volume with lifetime dependent efficiency. Furthermore, 
the decay branching ratio of $\gamma'\to e^+e^-$ is comparable with that of $\gamma'\to a \gamma$ when $\varepsilon \lesssim 10^{-6}$ for $m_{\gamma'}\lesssim 0.7$ GeV. As a result, the mass range covered by beam-dump experiments becomes smaller.  If $g_{a\gamma\gamma'}$ is increased to $10^{-2}~\rm{GeV}^{-1}$, see the upper-right panel, the sensitivities of beam-dump experiments for $m_{\gamma'}$ heavier than $10$ MeV are lost with $\varepsilon \lesssim 10^{-4}$. 

For LHCb , BaBar and NA48/2, which are sensitive to the prompt decay of heavy dark photon with $\varepsilon \gtrsim 10^{-3}$, the branching of $\gamma'\to e^+e^-,\mu^{+}\mu^{-},\,\text{and}\,\text{hadrons}$ could be comparable with $\gamma'\to a \gamma$. Therefore, the lower bounds of $\varepsilon$ change slightly when $g_{a\gamma\gamma'}=10^{-2}\,\rm{GeV}^{-1}$, see upper-right panel of Fig.~~\ref{fig:limit}.
 In the case of large $g_{a\gamma\gamma'}$, e.g. $10^{-1}~\rm{GeV}^{-1}$, the decay of dark photon is dominated by $\gamma'\to a\gamma$ channel for $m_{\gamma'}\gtrsim 0.1$ GeV. Therefore, the lower limits of $\varepsilon$ from LHCb and BaBar searches become loose, especially for heavier dark photon, see lower-left panel of Fig.\ref{fig:limit}. The allowed $\varepsilon$ can even reach ${\cal O}(0.01)$ for $m_{\gamma'}\gtrsim 10$ GeV.  However, when we take into account the constraints on  $g_{a\gamma\gamma'}$, the parameter space for dark photon lighter than $10$ GeV  is inconsistent with the constraints of mono-photon searches at BaBar~\cite{deNiverville:2018hrc} shown by the vertical dashed line. If we take $g_{a\gamma\gamma'}$ to be larger, e.g. $1.0~{\rm GeV}^{-1}$, the LHCb constraints is getting more and more released, and a large parameter space is consistent with muon $g-2$ result, see the red band in the lower-right of Fig.~~\ref{fig:limit}. 
    
\section{Summary }
\label{sec:Summary}

The existence of dark sector with portal-type coupling provides a natural and simple way for new physics that weakly interacts with the SM. Through the $U(1)$ kinetic mixing term between dark sector and SM, a massive dark photon from dark sector could couple to SM fermions, and the deviation of muon $g-2$ from SM can be explained. However, experimental searches for dark photon impose stringent bounds on the mixing parameter $\varepsilon$ and exclude the possibility for dark photon to solve the muon $g-2$ anomaly. In this study, we propose that a light pseudo-Goldstone boson, axion-like particle, may exist in dark sector. Therefore, dark photon could decay into it with a photon, if this decay channel is kinematically allowed. Due to this additional decay channel, we found significant changes in the bounds on $\varepsilon$ imposed by both beam-dump and collider experiments. When the coupling $g_{a\gamma\gamma'}$ between axion-like particle, dark photon and photon reaches $1.0~{\rm GeV}^{-1}$, the dominant decay channel of dark photon will be $\gamma'\to a \gamma$ for mass of dark photon heavier than $0.1$ GeV. As a result, the parameter space that is consistent with observation of muon $g-2$, but previously excluded by LHCb, is now allowed.

\section*{Acknowledgment}
\label{sec:Acknowledgment}
We would like to acknowledge the support of National Center for Theoretical Sciences (NCTS). This work was supported in part by the National Science and Technology Council (NSTC) of Taiwan under Grant No.MOST 111-2112-M-003-006, 111-2811-M-003-025-, 112-2811-M-003-004, and NSTC 113-2811-M-003-019.

\end{document}